\documentclass[useAMS,usenatbib,usegraphicx]{mn2e}

\title[Formation of Kuiper-belt binaries]
{Formation of Kuiper-belt binaries through multiple chaotic
scattering encounters with low-mass intruders}
\author[Astakhov, Lee and Farrelly]{Sergey A.~Astakhov,$^{1}$\thanks{E-mail:
s.astakhov@fz-juelich.de; www.astakhov.newmail.ru} Ernestine A.~
Lee $^{2,3}$\thanks{E-mail: ernestine.lee@gmail.com} and David
Farrelly$^{2}$\thanks{E-mail: david.farrelly@gmail.com}
\\
$^{1}$John von Neumann Institute for Computing, Forschungszentrum J\"ulich, D-52425 J\"ulich, Germany\\
$^{2}$Department of Chemistry and Biochemistry, Utah State
University, Logan, UT 84322-0300, USA \\
$^{3}$FivePrime Therapeutics, 951 Gateway Boulevard, South San
Francisco, CA 94080, USA}

\begin{document}

\date{Submitted 2004 November 7; Accepted 2005 April 4}

\maketitle

\begin{abstract}
The discovery that many trans-neptunian objects exist in pairs, or binaries,
is proving invaluable for shedding light on the formation, evolution
and structure of the outer Solar system. Based on recent systematic searches it has been
estimated that up to 10\% of Kuiper-belt objects might be binaries.
However, all examples discovered to-date are unusual, as compared to
near-Earth and main-belt asteroid binaries, for their mass ratios of
order unity and their large, eccentric orbits. In this article we
propose a common dynamical origin for these compositional and
orbital properties based on four-body simulations in the Hill
approximation. Our calculations suggest that binaries are produced
through the following chain of events: initially, long-lived
quasi-bound binaries form by two bodies getting entangled in thin
layers of dynamical chaos produced by solar tides within the Hill
sphere. Next, energy transfer through gravitational scattering with
a low-mass intruder nudges the binary into a nearby non-chaotic,
stable zone of phase space. Finally, the binary hardens (loses
energy) through a series of relatively gentle
gravitational scattering encounters with further intruders. This
produces binary orbits that are well fitted by Kepler ellipses.
Dynamically, the overall process is strongly favored if the original
quasi-bound binary contains comparable masses. We propose a
simplified model of chaotic scattering to explain these results. Our
findings suggest that the observed preference for roughly equal mass
ratio binaries is probably a real effect;  that is, it is not
primarily due to an observational bias for widely separated,
comparably bright objects. Nevertheless, we predict that a sizeable
population of very unequal mass Kuiper-belt binaries is likely
awaiting discovery.
\end{abstract}


\section{Introduction}

Binary systems occupy a special place in astronomy. First and
foremost they allow the determination of binary partner masses
which, in the case of stars, proved critical in the early
development of the mass-luminosity relation \citep{eddington,noll}.
In fact, the majority of all stars seem to be members of
gravitationally bound multiplets with most being members of binaries
\citep{heggie}. The dynamics and energetics of star clusters, in
particular, are strongly influenced by the presence of binaries.
This is because the largest single contributor to the total
mechanical energy of the cluster may be the internal energy (binding
energy) of binaries \citep{hills0, inagaki,janes,heggie}. Not
surprisingly the dynamics of binary systems and their formation
mechanism are a subject of continuing interest in stellar
dynamics \citep{heg, hills0, hills1, hills2, hills3, mikkola,
mard1,mard2,quinlana,heggie}.

\begin{table*}
  \centering
  \begin{tabular}{p{5cm}p{5cm}p{5cm}}
  \hline
   \textbf{Reference} & \textbf{Mechanism} & \textbf{Notes} \\\\

  Path 1 \citep{weiden} &

  Physical collision and subsequent accretion of two bodies inside Hill
  sphere of a third, larger Kuiper-belt object. &

   Depends on existence of approximately two orders of magnitude more massive bodies in the primordial
Kuiper-belt than is currently accepted. At time of binary formation 99\% of total mass of Kuiper-belt
thought to be in much smaller bodies \citep{gold, kenyon}. \\\\

  Path 2 \citep{gold} &

  Transitory binary formation inside Sun-binary Hill sphere.
  Capture and stabilization by dynamical friction. &

  Predicts primordial  binary formation rate
  $\approx 3 \cdot 10^{-6}$ per year per large body. Assumes larger sea of small ($<~10$~km) bodies than gravitational
   instability theories predict \citep{funato}. Predicts formation of multiplets.
  \\\\

  Path 3 \citep{gold} &
  Transitory binary formation inside Sun-binary Hill sphere.
  Capture and stabilization by gravitational scattering with
  a single large intruder -- $L^3$ mechanism.&

   Predicts primordial binary formation rate $\approx 3 \cdot 10^{-7}$ per year per large body.
   No explicit mechanism provided for gradual reduction in post-capture binary orbit
   semimajor axes. \\\\

  Path 4 \citep{funato} &
  Initial production of asymmetric mass ratio ``asteroid-like'' binaries through physical collisions.
  Subsequent scattering encounters with large intruders then equalize mass ratio through
  exchange reactions.&

  Neglects solar tidal effects. Exclusively produces binaries with eccentricities $e~>~0.8$.
  Effect of varying intruder mass not studied. Predicts future TNO binary discoveries
  will have roughly equal masses, large separations and high eccentricities.\\\\

  Chaos-assisted capture (this article) &
  Transitory binary formation through chaos-assisted capture \citep{cac}
  inside Sun-binary Hill sphere. Subsequent stabilization through a sequence
  of gravitational scattering encounters with small intuders each having ca. $1-2\%$ of total binary mass. &

  Includes solar tidal effects. Predicts {\it ab initio} the formation of (i) binaries with moderate eccentricities $0.2 <  ~e~  < 0.8$ and (ii)
  a preference for roughly equal mass binaries. Also predicts that
  future higher resolution surveys of the Kuiper-belt will discover populations of asymmetric mass binary
  TNOs. Mechanism suggests that binaries formed when the KB disk was dynamically cold, {\it i.e.,}
  prior to excitation and depletion of the Kuiper-belt (\citealt{dunc, gomes}; \citealt*{lev}).\\\hline

 \end{tabular}
  \caption {Proposed Kuiper-belt binary formation mechanisms.} \label{tbl}
\end{table*}

Binaries are also relatively common in planetary physics; these
include the Earth-Moon and Pluto-Charon systems \citep{chr, canup, canupa,
canup1} as well as a substantial number of asteroid binaries in the
main belt. Asteroid binaries usually contain partners with mass
ratios in the range $m_r \sim 10^{-3}-10^{-4}$ where $m_r =
m_{2}/m_{1}$ and $m_1$ and $m_2$ are the masses of the primary and
secondary binary partners respectively. Objects with such asymmetric masses
are most often thought of as ``asteroids with satellites''
\citep{merline} rather than as ``binaries'' although there exists no
meaningful quantitative distinction between these definitions. The
recent discovery of binary trans-neptunian objects (TNOs) is
particularly significant because these objects allow for the direct
determination of TNO masses thereby opening up an important window
into the origin, evolution and current composition of the
Kuiper-belt (KB) \citep{stewart,toth, luu, noll, glad}. It is quite remarkable that recent systematic searches \citep{iau1, noll, schaller} seem to suggest that up to 10\% of Kuiper-belt objects might exist as gravitationally bound pairs \citep{burns}.

Kuiper-belt binaries (KBBs) are notable for their unusual
compositional and orbital properties as compared to main-belt
asteroids: the most salient
 are \citep{noll} $(i)$ large
mutual orbits, $(ii)$ highly eccentric mutual orbits and $(iii)$
mass ratios of order unity (\citealt{margot, veillet, durda,
weiden}; \citealt*{gold}; \citealt{noll, schaller};
\citealt*{osip,alt}; \citealt{funato,burns, noll1,noll2,tak,naz}).
However, in each case certain caveats apply
\citep{noll,noll1,noll2}. Firstly, by some measures KBB orbits are
indeed large; {\it e.g.}, for main-belt asteroids with satellites
the semimajor axis of the mutual orbit ($a$)\ is typically about an
order of magnitude larger than the radius of the primary asteroid
($r_A$) whereas for KBBs $a/r_A \sim 50-500$ \citep{merline,noll2}.
Curiously, however, if
the semimajor axis is compared to the radius of the binary's Hill sphere ($%
R_{H}$) then main-belt asteroid binaries and KBBs have rather
similar semimajor axes, {\it i.e.}, expressed as a percentage $a/R_H
\sim 2-5\%$ \citep{noll,noll2}. Secondly, although KBB orbits are
noticeably eccentric as compared to the almost circular orbits of
main-belt asteroid binaries, recent observations suggest that extremely large eccentricities
($e>0.8$) may be rarer than was initially thought \citep{noll,noll2}. Finally,
while the mass ratios of known KBBs are in the range $m_r \sim
0.1-1$ it is also possible that this result is at least partially
due to an observational bias for well-separated, equally bright
objects \citep{burns}. Nevertheless, the properties of KBBs are
sufficiently striking to suggest that their formation mechanism
differed considerably from that of main-belt or near-Earth asteroid
binaries, which are thought to have been produced by physical
collisions (\citealt*{chauvineau}; \citealt{xu}; \citealt{dur}).

Four different pathways \citep{weiden,gold,funato} -- summarized in
Table~\ref{tbl} -- have been proposed to explain the formation and
properties of KBBs which we will refer to as Paths 1 -- 4: Path~1
involves physical collisions between two planetesimals within the
Hill sphere of a larger Kuiper-belt object \citep{weiden}. The two
bodies then accrete and remain gravitationally bound as a single
object to the larger body, thereby producing a binary. This
mechanism assumes that binary objects are primordial; Similar to
Weidenschilling's mechanism \citep{weiden} Paths 2 and 3 also
proceed from the temporary gravitational capture of two objects
within the three-body Hill sphere. Now, however, the large third
body is the Sun \citep{gold}. In both Paths 2 and 3 the initial
stage of binary formation is the formation of a transitory binary
which will eventually ionize unless stabilization occurs first. Two
stabilization mechanisms have been proposed; in Path~2 stabilization
results from dynamical friction through interactions with a sea of
smaller ($< 10$ km) Kuiper-belt bodies. In Path~3 the transitory
binary is stabilized by energy loss through gravitational scattering
with a large third body -- the $L^3$ channel. Both of these
mechanisms emphasize the importance of the Sun-binary Hill sphere
\citep{sze, murray}.

A common feature of Paths 2 and 3 is that immediately before and
after capture the binary partners are following an essentially
three-body orbit. That is, solar perturbations or tides are
important. Thus the orbits are periodic, quasi-periodic or chaotic
\citep{ll} and, in general, cannot be well described by orbital
elements of the Kepler problem \citep{murray}. However, actual KBBs
follow orbits which can be well fitted by Kepler ellipses
\citep{veillet,noll,osip,noll1,noll2}, {\it i.e.}, solar tides are
relatively unimportant and the orbits are approximately two-body in
nature. Therefore, a mechanism for gradually transforming captured
-- and, therefore, bound -- quasiperiodic three-body orbits into
(almost) periodic two-body orbits is required. We term this general process
``Keplerization'' -- {\it i.e.}, the gradual energy loss of
three-body binary orbits to produce orbits that are essentially
Kepler ellipses. As three-body orbits lose energy their semimajor
axes undergo a steady reduction in size. In principle, Keplerization
may occur directly during the capture process itself ({\it e.g.}, as
posited to happen in $L^3$) or it can happen more gradually after
the initial capture event through, {\it e.g.}, continued dynamical
friction over extended timescales \citep{gold}.

Path~4, proposed by \citet{funato}, involves exchange reactions
\citep{heggie} in which the smaller member of an already bound
``asteroid-like'' binary is displaced by a larger body during a
three-body encounter. This mechanism is essentially a variant of
binary star-intruder scattering
\citep{hills0,hills1,hills2,hills3,hut,heggie}. In this mechanism
solar tides are ignored and both the pre- and post-collision binary
(if it survives) follow a two-body Keplerian ellipse.

While each of these four pathways is feasible, no consensus has yet
emerged as to their relative importance or even whether a different
mechanism altogether might have operated \citep{burns}. This is
because each mechanism admits at least one potentially serious
drawback. The first three paths are sensitive to assumptions about
the size distributions of Kuiper-belt objects \citep{funato}. Path~1
depends on the existence of more large bodies in the Kuiper-belt
than seems to be consistent with observations \citep{gold, funato};
The basis of Path~2 is dynamical friction which, to be effective,
requires a much larger sea of small bodies than is predicted by
current theories of planetesimal formation \citep{gw, wetherhill,
rafikov, bern, funato, kenyon}. Further, it is unclear how and why
dynamical friction should select preferentially for roughly equal
mass ratio binaries. This assumes, of course, that the apparent
preference for mass ratios of order unity is not an observational
artefact \citep{burns}.

The $L^{3}$ channel \citep{gold}, Path~3, depends on relatively rare
close encounters between three large objects and, based on the
estimates of  \citet{gold}, it is unclear if these occurred often
enough to produce the current population of KBBs. In this regard it
is interesting to note that recent calculations suggest that, if the
 Kuiper-belt lost its mass through collisional grinding,
then an order of magnitude more binaries were likely present in the
primordial Kuiper-belt than has otherwise been thought
\citep{petit}. This illustrates that the number and size
distributions of primordial KB objects might be subject to revision
\citep{stern, bern, kenyon, petit, ell}. A second drawback to Path~3
is that, as originally proposed, it provides no explicit mechanism
-- beyond possible post-encounter dynamical friction -- for
producing the approximately two-body Keplerian binary orbits which
are actually observed \citep{veillet, noll, schaller, osip, noll1,
noll2}. It is unlikely, as our simulations confirm, that a single
collision between a quasi-bound binary ({\it i.e., } one that is
following a three-body Hill trajectory) and a massive intruder will
result in a two-body Keplerian binary orbit. Finally, Path~4 leads
almost exclusively to binary eccentricities $e \geq 0.8$ and very
large semimajor axes which may approach $R_{H}$ itself
\citep{funato}. However, moderate eccentricities, in the range $0.25
\leq e \leq 0.82$, and semimajor axes that are only a few percent of
$R_H$ seem to be the rule \citep{noll, osip, noll1, noll2}.

In the model described herein the first step is the formation of
long-lived quasi-bound binaries through the recently proposed
mechanism of chaos-assisted capture \citep[CAC,][]{cac,af,trimble}
in the Hill approximation \citep{sze, henon, murray, heggie}. In CAC
particles become temporarily caught-up in thin chaotic layers within
the Hill sphere of, in this case, the Sun-binary system. These chaotic layers,
which separate regular from asymptotically hyperbolic (scattering) orbits, are
the result of the perturbation of the Sun on the motion of the two
bodies making up the binary. Because the orbits are chaotic any
quasi-bound (transitory) binary which is formed will eventually
break apart. However, the lifetimes of these quasi-bound objects can
be sufficiently long that, in the interim, stabilization can occur.
Since the chaotic layers in the Hill problem \citep{simo, cac, af}
lie adjacent to regular Kolmogorov-Arnold-Moser
\citep[KAM,][]{zas,ll} regions, then, in principle, even relatively
weak perturbations can switch the binary into such KAM regions and thereby lead to permanent capture.

Of course, the precise nature of the stabilization mechanism is the
crux of the matter. We propose that capture and subsequent hardening
of three-body Hill orbits ({\it i.e.,} Keplerization) \citep{hills0,
hills1,hills2,hills3,quinlana,heggie} proceed through chaotic
gravitational scattering encounters with low-mass intruders which
happen to transit the Hill sphere. Once a binary has been captured
initially then further encounters with intruders
are probable -- about once every 3,000-10,000 years \citep{weiden}
-- and this can eventually produce binaries whose mutual orbits are
Kepler ellipses. Empirically we find that the process is most
efficient for equal mass binaries and relatively low-mass intruders
having $\sim 2\%$ or less of the total binary mass. Incidentally, we
 argue that CAC is also the dynamical basis of the twin
mechanisms proposed empirically by \citet{gold}.

To model the statistics of capture we compute capture probabilities
in four-body (Sun, binary, intruder) Monte Carlo scattering
simulations for various binary mass ratios in the Hill approximation
\citep{sch}. Our simulations reveal a clear preference for equal
mass binaries to survive multiple subsequent encounters as they
harden through intruder scattering.

The paper is organized as follows; Section~\ref{sec2} introduces the
equations of motion for the three- and four-body problem in the Hill
approximation and provides a brief overview of how chaos can assist
capture into stable three-body orbits \citep{cac, af}. Monte Carlo
simulations in the four-body Hill approximation \citep{sch} are
described in Sec.~\ref{sec3}. A simplified (reduced dimensionality)
model of chaotic low-mass intruder scattering is introduced in
Sec.~\ref{sec4} to explain our results. To study the dynamics in
this model we employ the Fast Lyapunov Indicator for several mass
ratio and binary eccentricity combinations. A more detailed
comparison with the other formation models summarized in
Table~\ref{tbl} is made in Sec.~\ref{sec5}. Conclusions are in
Sec.~\ref{sec6}.

\section{Equations of motion in the three-body and four-body Hill problems}
\label{sec2}

 In our model the initial stages of binary formation
involve temporary capture within the Sun-binary Hill sphere through
two particles getting entangled in chaotic regions of phase space,
{\it i.e.}, chaos-assisted capture \citep{cac, af}. This process is
well described in the three-body Hill approximation.

\subsection {Three-body Hill problem}

The three-body vector equations of motion in the Hill approximation
are given by \citep{sze, henon,murray,sch}

\begin{equation}
\ddot{\brho}+\mathbf{\Omega }\times [ 2\dot{\brho}+{\bf \Omega}
\times \brho]=-\brho+3 {\bf a}({\bf a} \cdot
\brho)-\frac{\brho}{|\brho|^{3}} \label{hill}
\end{equation}

\noindent{Here} $\brho={\brho}_{2}-{\brho}_{1}$ is the relative
distance between the binary members which have masses $m_{1}$ and
$m_{2}$ and whose centre-of-mass is assumed to move along a
circular orbit $\mathbf{a}=(1,0,0)$ around a larger body $m_0$
(the Sun). The coordinate system $(x,y,z)$ is rotating with constant
angular velocity ${\bf \Omega}=(0,0,1)$ which corresponds (in scaled units)
to the mean motion of the binary centre of mass. In this coordinate system
the centre-of-mass of the binary is located at the origin as is usual in the Hill problem \citep{murray}.

\begin{figure}
\begin{center}
\includegraphics{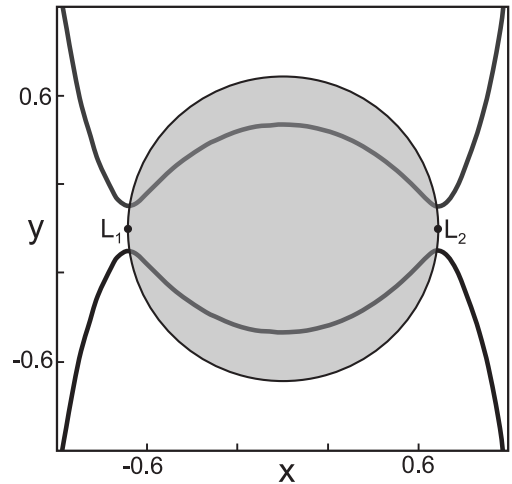}
\caption{\label{fig1} Typical level curves of the zero-velocity
surface together with the Lagrange points $L_1$ and $L_2$ for the
three-body Hill problem in the planar limit \citep{murray}. The
circle shaded grey shows the location of the Hill sphere. The
Lagrange points demark the gateways into the capture zone. All units
are scaled Hill units.}
\end{center}
\end{figure}

Units and typical values of parameters are defined as
follows with reference to $1998~WW_{31}$ \citep{veillet}: $m_{1}\sim 2\cdot 10^{21}g$, $%
m_{2}\sim 10^{21}g$; the semimajor axis of the primary orbit
$a_{b}\sim 45AU$ and of the mutual binary orbit $a\sim 22000$ km.
The Hill sphere (see Fig.~\ref{fig1}) occupies the region between
the saddle points $L_1$ and $L_2$ of the Hill problem which we term
the ``capture zone'' \citep{sze, sch, murray}; The Hill radius
$R_{H}=a_{b}\left( \frac{m_{1}+m_{2}}{3m_{0}}\right) ^{1/3}$; in
physical units: $R_{H}\sim 3\cdot 10^{5}$km~$\sim 25a$. In scaled
Hill units \citep{sze, murray} $R_{H}=(1/3)^{1/3}$ and the radii of
the binary partners $r_{A}\sim 10^{-4}$.

\subsection{Chaos-assisted capture in Hill's problem}

\begin{figure*}
\begin{center}
\includegraphics{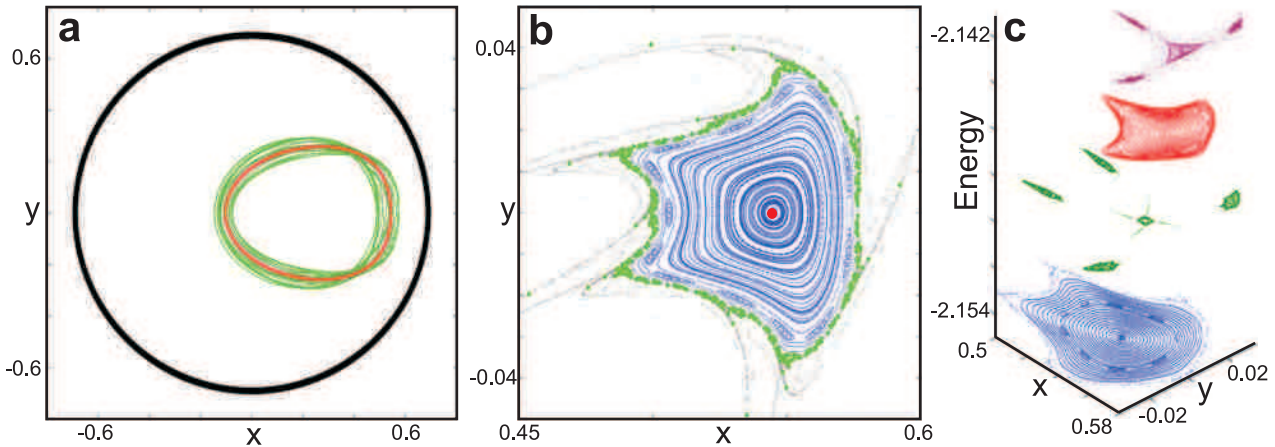}
\caption{\label{fig2} Non-linear dynamics of transitory binary
formation in Hill's three-body problem. Panel (a) shows a
quasi-bound (chaotic) binary centre-of-mass orbit (green)
temporarily trapped close to a periodic orbit (red) which lies at
the centre of a stable KAM island. This trajectory eventually
escapes from the Sun-binary Hill sphere (large black circle). Panel
(b) is a composite showing the colour coded SOS ($x-y$ hyperplane
defined such that the momentum $p_x=0$ and the velocity ${dy \over
dt} > 0$) for the trajectories in panel (a) and also several nearby
chaotic (grey) and quasi-periodic (blue) orbits. Panel (c) shows the
non-linear evolution of the size and shape of the KAM island in (b)
with energy at four selected energy values. Distances and energies
are in scaled Hill units.}
\end{center}
\end{figure*}

In order for a transitory (quasi-bound) binary to form two bodies
must come inside their mutual Hill sphere defined with respect to
the Sun. The Lagrange saddle points $L_1$ and $L_2$ in
Fig.~\ref{fig1} serve as gateways between the interior of the Hill
sphere and heliocentric orbits. Figure \ref{fig2}a shows a typical
chaotic binary orbit obtained by integrating Hill's equations
(\ref{hill}) in two-dimensions. The orbit is trapped close to a
periodic orbit for many periods before finally escaping from the
Hill sphere (not shown). The accompanying Poincar\'{e} surface of
section (SOS) in Fig.~\ref{fig2}b shows that the orbit is actually
trapped in a chaotic layer separating a regular KAM island from a
large region of hyperbolic scattering. Note that the orbits shown in
Fig.~\ref{fig2} correspond to the mutual Hill orbit of the two
binary partners and are independent of the particular mass ratio of
the binary partners \citep{henon, murray}.

Examination of
Poincar\'{e} surfaces of section in the Hill problem
\citep{simo} -- or, equivalently, the circular restricted three-body
problem (CRTBP) for small masses \citep{cac} --
reveals that, at energies above the Lagrange saddle points $L_1$ and $L_2$ all phase space is
divided into three parts, one of which regular KAM orbits inhabit, chaotic orbits another,
the third consists of direct scattering orbits or, in other words, hyperbolic orbits.
All these differ from each other in the behaviour of their orbits.
The chaotic orbits separate the regular from the hyperbolic regions -- see Fig.~\ref{fig2}b. As
energy is increased above the saddle points the chaotic regions
consist of relatively thin layers of chaos which cling to the progressively
shrinking KAM tori \citep{simo,cac,af}. Because incoming orbits
cannot penetrate the regular KAM tori in 2-dimensions (2D) and
only exponentially slowly in 3D \citep{ll,cac}, orbits entering
the Hill sphere from heliocentric orbits must either enter chaotic
layers or scatter out of the Hill sphere essentially immediately.

Binaries corresponding to chaotic orbits lying close to KAM tori can
themselves follow almost regular orbits for very long times as shown in
Fig.~\ref{fig2}a. In the presence of a source of dissipation or of
other perturbations these orbits can easily be switched into the nearby, but otherwise impenetrable, KAM
regions.

It is remarkable that capture can occur if the binary decreases or
increases its energy during the encounter. This can happen because
the topologies of stable KAM islands are, in general, a non-linear
function of energy as is illustrated for four selected energies in
Fig.~\ref{fig2}c. That is, if a perturbation causes a chaotic orbit
to increase or decrease its energy suddenly it may, in either event,
find itself captured in a nearby regular part of phase space.
Alternatively, such perturbations, though small, may cause the
quasi-bound binary to break up more quickly if the size of the
chaotic layer is suddenly reduced in favour of scattering regions.

Possible energy loss mechanisms which can stabilize transitory
binaries include physical collisions \citep{weiden}, gravitational
scattering with other bodies and dynamical friction \citep{gold}.
Ruling out physical collisions and dynamical friction for reasons
already described leaves gravitational scattering with
``intruder'' objects which enter the binary Hill sphere. The efficiency of this process depends
critically on the intruder mass. For example, we find that the
delicate chaotic binary orbits tend to be catastrophically destabilized
by large intruders as, {\it e.g.}, in the $L^3$ mechanism
\citep{gold}. Further, as Fig.~\ref{fig2} shows, the orbits of
transitory binaries can be a very large fraction of the Hill radius
whereas actual binary orbits are typically only a few percent of
$R_H$. Any KBB production mechanism must, therefore, explain how
these large initial orbits are transmogrified into compact,
Keplerian orbits. It is unlikely (as our simulations will show) that
a single scattering encounter would propel the binary into such a
final state directly. Rather, we propose that a series of gentle
 scattering encounters with low-mass intruders is necessary to capture and finally
produce Keplerian orbits. Gentle
stabilization by low mass intruder scattering is possible because
only a weak perturbation is needed to drive long-lived chaotic
orbits into adjacent regular regions of phase space.

\subsection{Four-body Hill problem}

The simulation of intruder scattering is facilitated by studying
intruder scattering in the four-body (Sun, binary, intruder) Hill
approximation \citep{sch, brasser, sch1}. This approximation is
appropriate given the mass ratios involved and the low to moderate
eccentricities of typical KBB centre-of-mass orbits around the Sun.

The generalization of the Hill problem \citep{sze,henon, murray}
to the case where three small bodies, with a mutual
centre-of-mass, ${\bf R}_c$, orbit a much larger fourth body (the
Sun, $m_0=1$) on a near circular orbit ({\it e.g.}, three
Kuiper-belt objects within their mutual Hill sphere) is due to
\citet{sch}. The total mass of the three bodies is defined by

\begin{equation}
\mu =\sum_{j=1}^3 m_j \ll 1  \label{mu}
\end{equation}

\noindent where ${\bf R}_c \approx {\bf a}=(1,0,0)$ defines the
motion of the three-body centre-of-mass along an almost circular
orbit which defines the rotating frame. The vector equations of motion are \citep{sch}

\[
\ddot{\brho} + {\bf\Omega} \times [2 \dot{\brho} + {\bf\Omega}
\times \brho]=-\brho+3{\bf a}({\bf a} \cdot \brho) - ({\alpha}_1 +
{\alpha}_2) \frac {\brho} {|\brho|^3} +
\]

\begin{equation}
 {\alpha}_3 \left( \frac {{\brho}_3 -{\brho}_2}{|{\brho}_3 -
{\brho}_2|^3} -  \frac {{\brho}_3 -{\brho}_1}{|{\brho}_3 -
{\brho}_1|^3}\right)
  \label{rho}
\end{equation}

\[
\ddot{\brho}_3 + {\bf\Omega} \times [2 \dot{\brho}_3 + {\bf\Omega}
\times {\brho}_3]=-{\brho}_3+3{\bf a}({\bf a} \cdot {\brho}_3) -
\]

\begin{equation}
 {\alpha}_1  \frac {{\brho}_3 -{\brho}_1}{|{\brho}_3 -
{\brho}_1|^3} -  {\alpha}_2 \frac {{\brho}_3
-{\brho}_2}{|{\brho}_3 - {\brho}_2|^3}
   \label{rho3}
\end{equation}

Here ${\brho}_3$ is the coordinate of the third intruder body, $m_3$, and
$m_{j}=\mu {\alpha }_{j}$ where

\begin{equation}
\sum_{j=1}^3 {\alpha}_j = 1.  \label{alpha}
\end{equation}

Equations (\ref{rho}) and (\ref{rho3}) contain the coordinates
${\brho}_1, {\brho}_2$ explicitly; this is purely for compactness
of notation and, in practice, they are eliminated using the
centre-of-mass relation

\begin{equation}
\sum_{j=1}^3 {\alpha}_j {\brho}_j= 0.  \label{com}
\end{equation}

When $m_{3}=0$ equation (\ref{rho}) reduces to the three-body Hill
problem (\ref{hill}) \citep{sze,murray} and becomes uncoupled from
equation (\ref{rho3}). Equation (\ref{rho3}) is then a variant of
the restricted Hill four-body problem \citep{sch} and depends
explicitly on the solution $\brho(t)$ which may be periodic,
quasi-periodic, or chaotic \citep{ll,cac}. These solutions are
actually particular trajectories for the three-body Hill problem.

To model the statistics of capture immediately after transitory
binary formation we computed capture probabilities in four-body
Monte Carlo scattering simulations as described in the next
section.

\section{Monte Carlo Scattering Simulations}
\label{sec3}
 We are primarily interested in binary-intruder scattering, with
intruders approaching from infinity. Therefore we assume that the
transitory binary is initially following a chaotic, but long-lived
{\it three-body} Hill orbit.

Because our Monte Carlo scattering simulations involve chaotic
binary orbits they differ in essential respects from previous
simulations; to our knowledge all previous simulations have dealt
with intruder scattering from bound binaries which follow circular
or elliptical two-body Keplerian orbits \citep{hills0, hills1,
hills2, hills3,hut, heggie, funato}. Therefore it is necessary to
explain our procedure in detail.

In any energy range for which chaotic layers exist it is possible to
find an infinite number of initial conditions corresponding to
different quasi-bound, three-body binary orbits \citep{ll}. This is
illustrated in Fig.~\ref{fig2}c; it is apparent that a swath of
chaotic orbits over the energy range shown will not only have a
range of lifetimes but their dynamics will depend sensitively on
their particular initial conditions and energy. Since we are interested in the
probability of capture from chaotic zones into stable KAM zones it
is, therefore, necessary to study an {\it ensemble} of such orbits
rather than a single example. That is, the capture probability of
interest is not simply the probability of any given orbit being
captured; it must also measure the density of ``capturable'' orbits
inhabiting the chaotic zones.

\subsection{Methods}

The procedure we have developed is as follows:

\begin{enumerate}
\item Initially a cohort of 170 different quasi-bound chaotic binary orbit initial
conditions was harvested by integrating eq. (\ref{hill})  using a
Bulirsch-Stoer method \citep{NR,aarseth}. Initial conditions were
chosen randomly inside the three-body (Sun-binary) Hill sphere for
randomly chosen negative energies lying above the (three-body Hill)
Lagrange saddle points and in the energy range $-2.15<E<-2.0.$ If an
orbit remained inside the Hill sphere for a time $T_{1}<T_{H}<T_{2}$
then its initial conditions were stored. We refer to $T_{H}$ as the
orbit's ``natural'' Hill lifetime. In scaled units we chose $T_{1}$
$=15$ ($\sim $ 600 years) and $T_{2}$ = 50. While much longer (and
shorter) living orbits are possible selected integrations for such
orbits produce comparable results. This procedure guarantees that
all orbits in the cohort are chaotic since, absent perturbations,
they all eventually exit the Hill sphere.

\item Equations (\ref{rho}) and (\ref{rho3}) were
integrated for various values of $m_r$ and $m_{3}$ using the stored initial
conditions for the binary, but rescaled to the actual values of
$m_{1}$ and $m_{2}$ used. Thus, the same cohort of three-body
quasi-bound Hill orbits was used for all mass ratios.

\item Intruder particles were fired at each binary orbit with the phase of the
binary chosen randomly. For each binary orbit 5000 intruders were sent in and, each time,
it was recorded whether the binary was stabilized or not. For each binary orbit this produces
a capture probability, {\it i.e.}, the fraction of the 5000 intruders that lead to stabilization
in a single encounter with a binary.

\item Intruders were started isotropically with uniformly chosen
random positions on a sphere of radius $R_{H}$, with uniform
random velocities $|v|\leq 5v_{H}$ where $ v_{H}$ is the Hill
velocity (\citealt{gold, funato}; \citealt*{ann})

\begin{equation}
v_H \sim \Omega R_H,  \label{vh}
\end{equation}

\noindent with $\Omega$ being the orbital frequency of the third
body around the binary at distance $R_H$.

\item In order to choose the phase of the binary randomly intruder integrations were started
at randomly chosen times $t \in(0,T_{H})$ along the binary orbit. In
practice this has the effect of sending the intruder towards the
binary at different relative configurations of the binary partners in phase space.

\item The integrations proceeded until either the binary broke up
or it survived for a time $t=10~T_{H}$ at which point it was
considered captured. After the intruder escaped from a sphere of
radius $3R_{H}$ the integrations for the binary continued assuming
$m_{3}=0$, {\it i.e.}, the usual three-body Hill equations
(\ref{hill}) were used.

\item At the end of the integration each captured binary and its
intruder were back integrated in time to ensure that the
binary (in the four-body integration) broke up in the past. This
helps protect against accidentally starting the binary-intruder
trinary in already bound regions of phase space.

\item  Integrations were stopped if particles came within a
distance $r_{A}=10^{-4}$ of each other.

\item These simulations were repeated ten times so as to be able to compute variances (error bars)
for the capture probabilities. Each time a different random
number seed was used to guarantee that the simulations were
independent. Thus, {\it e.g.}, Orbit 1 may end up with capture probability
$0.4 \pm 0.02$, Orbit 2, $0.23 \pm 0.018$, and so on.

\item Orbits in the cohort were then binned according to their assigned capture probabilities.  This
produced histograms showing the number-density of orbits in the cohort having capture
probabilities in the various specified intervals. Mainly the error bars were used (a) to establish that the calculations had converged and (b) to decide
on a reasonable granularity for the probability bins.

\item The actual number of binary orbits included in the cohort was chosen so as to provide acceptable
convergence given available computational resources. Convergence
 was judged acceptable when variances in capture probabilities were on the order of a few percent or less.

\end{enumerate}

\begin{figure}
\begin{center}
\includegraphics{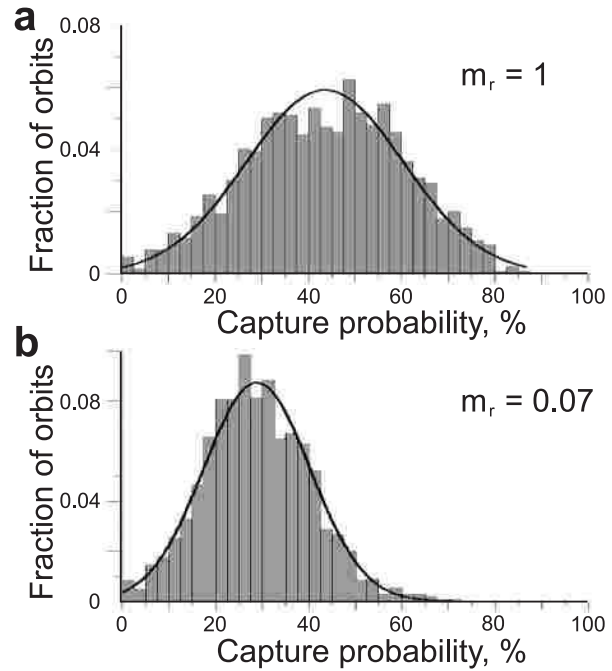}
\caption{\label{fig3} Fraction of binary orbits (number-density distributions)
binned according to their individual capture probabilities as
described in the text with ${\alpha}_3=0.025$; Mass ratios are
(a) ${\alpha}_1={\alpha}_2$ and (b) ${\alpha}_1=14{\alpha}_2$ (b).
Smooth curves are best fits to normal distributions. The same set of
170 chaotic Hill orbits in three-body centre-of-mass coordinates,
appropriately rescaled, was used in both cases. The binary was
considered captured only if the original binary partners remained
bound to each other, {\it i.e.}, exchange reactions were ignored. In
$98.5 \%$ of cases the average capture probability for any given
Hill orbit was higher for equal mass binaries.}
\end{center}
\end{figure}

\subsection{Single intruder scattering}

In these simulations, each binary orbit in the cohort ends
up with a capture probability and an error bar (of the order of a
few percent) attached to it. However, for different mass ratios
the same binary orbit in the cohort will, in general, have different capture
probabilities and error bars. Figure~\ref{fig3} presents histograms
showing the fraction of captured orbits binned according to
their individual capture probabililities for two typical mass
ratios. We tested that these histograms are robust to the actual
number of binary orbits included in our cohort -- {\it e.g.}, if 85
rather than 170 orbits are used the shapes of the histograms are
essentially identical to those in Fig.~\ref{fig3}. The width of the
bins is roughly commensurate with the average width of the error
bars.

The distributions in Fig.~\ref{fig3} clearly indicate that the
fraction of binary orbits peaks at a significantly higher capture
probability for equal mass ratio binaries than for unequal mass
ratio binaries. This translates into a higher probability for
capture in the equal mass case. Recall that the actual binary orbits
used in both cases are identical. Since the cohort of binary orbits
used is the same, and the intruder masses and initial conditions are
also the same, then the difference in the distributions is due
entirely to the different binary mass ratios employed. In fact the only
parameter that was varied between the simulations shown in Fig.~\ref{fig3} was the mass ratio.

We also
compared individual capture probabilities orbit for orbit and found
that in $98.5 \%$ of cases the average capture probability for any
given Hill orbit was higher in the equal mass simulation. Thus, at
any fixed Hill radius ({\it i.e.}, the same total binary mass) a
clear preference for the capture of same-sized binary partners is
apparent. For comparison, the mass ratios $m_r$ of the four best
characterized KBBs are $\sim 0.57$ ($1998~WW_{31}$,
\citealt{veillet}), $\sim 0.56$ ($(66652)~1999~RZ_{253}$,
\citealt{noll1}), $\sim 0.55$ ($(58534)~1997~CQ_{29}$,
\citealt{noll2}) and $\sim 0.34$ ($(88611)~2001~QT_{297}$,
\citealt{osip}). These are significantly larger than the mass ratios
of main-belt binaries for which, as noted, $m_r \sim
10^{-3}-10^{-4}$ \citep{merline}.

The calculations for Fig.~\ref{fig3} took $\sim 1$ month using all
nodes on a 32-node Beowulf cluster. Interestingly the mass effect
manifested itself quite directly in that the simulations for
Fig.~\ref{fig3}b finished about a week earlier than those for
Fig.~\ref{fig3}a. This is because of the additional tests and
integrations that needed to be performed to verify that an orbit had
been captured permanently.

\subsection{Multiple intruder scattering}

The probabilities in Fig.~\ref{fig3} are for permanently capturing
already-formed quasi-bound binaries through a {\it single} intruder
scattering event. However, the overall probability of binary
formation depends too on the probability of forming quasi-bound
binaries in the first place. Because Hill's equations (\ref{hill})
are parameter free \citep[after rescaling,][]{murray} once two
objects come within their mutual Hill sphere the probability of
chaos-assisted quasi-binary formation becomes independent of their
relative masses. Nevertheless, the entry rate of bodies into each
other's Hill sphere -- and thus the overall capture probability --
will depend on the actual masses and velocities of the objects
involved. Uncertainties in the size and velocity distributions of
contemporary and primordial Kuiper-belt objects means, however, that
estimating Hill sphere entry rates is an open problem but one whose
resolution should be aided by the discovery and characterization of
further KBB objects \citep{petit, bern, kenyon, noll2}. Despite
these uncertainties the mass effect is expected to persist because
it is progressively enhanced by later scattering events.

After its initial capture through a single scattering encounter with an intruder
a binary is, typically, following a very large three-body orbit
with a semimajor axis similar in size to that of the periodic orbit illustrated in Fig.~\ref{fig2}a.
That is, though now bound, the binary
partners are still strongly influenced by solar tides. In this subsection we
demonstrate that essentially Keplerian two-body orbits can be
produced through a series of subsequent scattering encounters with
further small intruders.

\begin{figure}
\begin{center}
\includegraphics{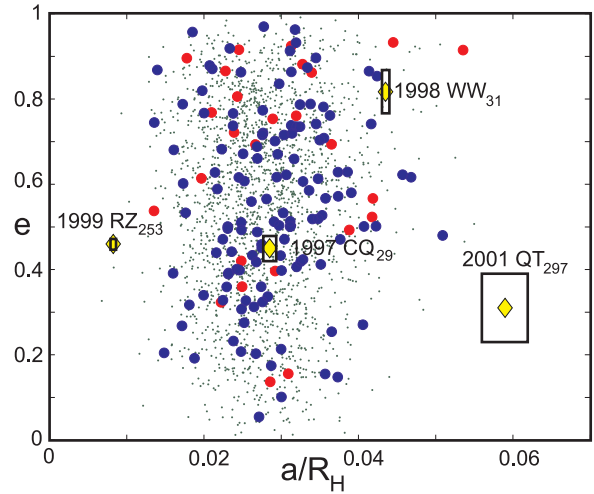}
\caption{\label{fig4} Distributions of orbital elements,
eccentricity, $e$, and semimajor axis, $a$ (scaled by the Hill radius
$R_H$) after $N_{enc} = 200$ intruder scattering events per binary
The figure combines results for twenty mass ratios:
$m_r=\frac{{\alpha}_2}{{\alpha}_1}=1$ (blue); $m_r=0.05$ (red) and
18 intermediate values $0.1 \leq m_r \leq 0.95$ (green dots). Yellow
diamonds show the locations of the four KBBs with known orbital
elements (boxes are $1 \sigma$ observational error bars)
\citep{veillet, osip, noll1, noll2}. For each mass ratio, 1200
different initial binaries were integrated to generate the figure.
To accumulate statistics this simulation was repeated 60 times for
the two extreme mass ratios. In each case 5000 (rather than 1200)
initial binaries were used per mass ratio. After $N_{enc}=200$
encounters binary survival probabilities were $0.103 \pm 0.007$
($m_r=1$) and $0.019 \pm 0.002$ ($m_r=0.05$). The survival
probability is defined as the fraction of the orginal 5000 binaries
which remains bound after $N_{enc}=200$.}
\end{center}
\end{figure}

That this is possible is demonstrated by the simulations underlying
Fig.~\ref{fig4}. In these simulations, for each mass ratio, a cohort
of 1200 randomly chosen Hill binary orbits (in the energy range $
-2.15 < E < -0.15$) was generated as described in the previous
subsection. The scatter plot shows the final orbital elements of
only those binaries out of the 1200 which survived $N_{enc} = 200$
successive encounters with intruders. Intruder initial conditions
were generated as described previously. In this set of
simulations an encounter was considered non-disruptive if the binary
survived for $t > 50$ scaled Hill time units. The first intruder was
sent in at a random point along the binary orbit during the orbit's
Hill lifetime $t \in (0,T_{H})$ (the binary orbit integration was
itself started at $t = 0$). If the binary survived to $t = 50$ the
next intruder was sent in. If the binary survived that event then
the next intruder was sent in, and so on. Thus, intruders were
spaced 50 time units apart. This was repeated until either the
binary broke up or it survived $N_{enc}$ such successive encounters.

In order to generate statistics it proved desirable to perform a
much larger simulation than was actually used to produce
Fig.~\ref{fig4}. The capture probabilities and error bars given in
the caption to Fig.~\ref{fig4} were obtained by repeating the
simulations as described in 60 independent runs, in each case using
an initial cohort of 5000 (instead of 1200) randomly chosen binary
orbits.

Naturally, each of the binary-intruder encounters may individually
be stabilizing or destabilizing. However, our simulations indicate
that comparable mass binaries, especially, tend to survive multiple
encounters as they harden through intruder scattering, thereby
becoming more Keplerian; this is similar to the situation in
low-mass intruder/binary-star scattering in which (depending on
intruder velocity) orbits tend, on average, to harden \citep{
hills0, hills1, hills2, hills3, heggie}. After $N_{enc} = 200$
intruder encounters Fig.~\ref{fig4} demonstrates that multiple
scattering can produce KBBs with orbital elements comparable to
those actually observed. Once a binary has been captured then
further encounters with intruders are likely. We assume $\sim 100$
km-sized binary partners and intruders having masses $~1-2 \%$ of the
total binary mass.Based on  Weidenschilling's model we find that
binary-intruder encounters will occur about once every 3,000-10,000
years \citep{weiden}; Fig.~\ref{fig4} demonstrates that such
encounters can eventually produce Keplerian orbits. Although the
final orbits in Fig.~\ref{fig4} are still strictly solutions to
Hill's problem, their energy has been reduced sufficiently that, for
all intents and purposes, they follow two-body Kepler ellipses, {\it
i.e.,} the orbits are well described by Keplerian orbital elements
\citep{murray}.

Experimental simulations for multiple intruder encounters reveal
that the main effect of increasing the number of encounters
($N_{enc}$) is to reduce the final value of the semimajor axis
whereas the final eccentricity distributions were quite robust to
$N_{enc}$. Thus we chose the single parameter $N_{enc} = 200$
heuristically so that orbits ended up roughly in the observed
semimajor axis range of $1998~WW_{31}$. Therefore, Fig.~\ref{fig4}
represents a single parameter fit to the semimajor axis of one of
the most well characterized KBBs. However, no special attempt was
made to refine the fit.

Kuiper-belt binary semimajor axes -- expressed as a fraction of the
binary Hill radius -- are generally comparable to those of asteroid
binaries and Fig.~\ref{fig4} captures this finding.
Figure~\ref{fig4} also reproduces the moderate eccentricities which
seem to be a feature of actual KBBs \citep{noll2}. This finding
contrasts with the extremely large eccentricities ($e > 0.8$)
produced exclusively in exchange reactions \citep{funato}. Finally,
Fig.~\ref{fig4} demonstrates that equal mass binaries have
significantly higher survival probabilities than asymmetric mass
binaries after multiple intruder encounters.

\section{Reduced Model of Intruder Scattering}
\label{sec4}

 The simulations described so far indicate that roughly
equal mass binaries have a statistically significantly higher capture
probability than binaries with small mass ratios. However, the
origin of this effect is not directly apparent from the simulations.
In this section we develop a reduced model of intruder scattering
which provides an explanation of the mass effect discovered in the
simulations. The origin of this effect is related to chaotic
scattering of the intruder by the binary.

\subsection{Chaotic scattering and dwell times}

Chaotic scattering involves a complex and sensitive dependence of
some ``output variable'' ({\it e.g.}, scattering angle, interplay-
or dwell-time, etc.) on an ``input variable'' ({\it e.g.}, impact
parameter). For example, scattering in the CRTBP has been
demonstrated to be chaotic \citep{boyd,benet,mac,nag}; \citet{benet}
plotted scattering functions (output energy and trajectory length)
as a function of impact parameter and found clear evidence of
chaotic scattering. Similar results were found by \citet{boyd} who
demonstrated chaotic scattering in binary star -- intruder
scattering. However, the high dimensionality of the four-body
problem, even in the Hill approximation, makes it difficult to study
a single output variable as a function of a single input variable.
Nevertheless, we performed a series of exploratory simulations in
which the dwell-time of the intruder inside the Hill sphere was
calculated as a function of binary mass ratio. The dwell-time is
here defined to be the total time the intruder remains inside the
Hill sphere in our Monte Carlo single-encounter scattering
simulations.

\begin{figure}
\begin{center}
\includegraphics{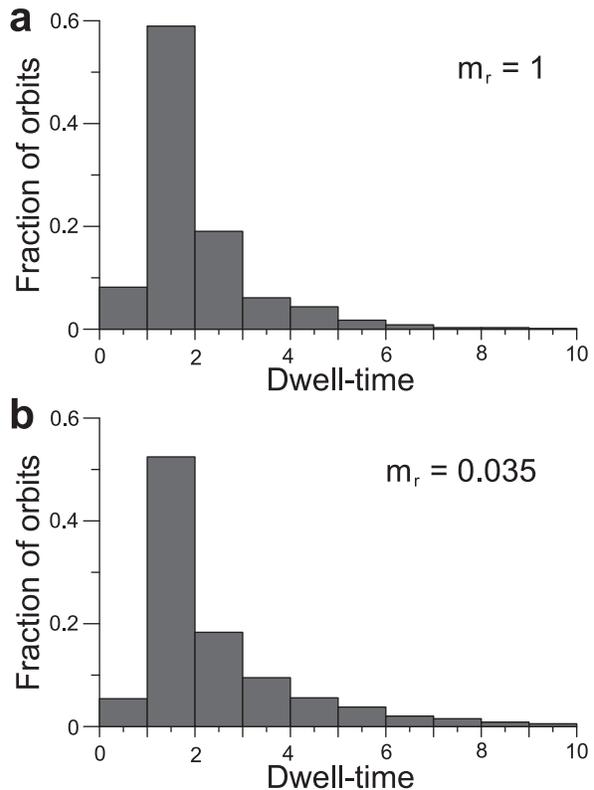}
\caption{\label{fig5} Dwell-time distributions for intruder
scattering for (a) $m_r=1$ and (b) $m_r = 0.035$. For each of the
two mass ratios shown 5000 low-mass intruder scattering events were
simulated as described in Sec.~\ref{sec3}. The histograms show
densities of intruders in various dwell-time ranges. Some
dwell-times $t > 10$ occur but these have been omitted from the
figure for clarity. For the sake of comparison, in these
simulations, a single chaotic binary orbit was used but the results
are similar if different chaotic binary orbits are used. An
identical series of intruders was used in both cases. The Hill
lifetime of the binary orbit used is $T_H \approx 17$ in scaled
units. Intruders are binned according to the time they remain within
the Hill sphere -- their dwell-time. Average dwell-times are (a) 2.1
and (b) 2.7 scaled time units.}
\end{center}
\end{figure}

Fig.~\ref{fig5} presents typical histograms of intruder dwell times
for two mass ratios. The dwell-time distributions for the two cases
differ in that for unequal masses the distribution has a longer tail
corresponding to longer dwell-times. That is, the intruder has a
greater chance of becoming caught up in a long-lived resonance \citep{heggie}
if the binary has very asymmetric mass partners. The average dwell
times shown are quite typical and are fairly consistent between
different binary orbits. The unequal mass case typically has an
average dwell-time $\approx 20-30\%$ longer than that for equal
masses depending on the actual mass ratio used and the particular
binary orbit selected.

In an attempt to understand this result and how (or if) it relates
to the differential stabilization of equal and unequal mass binaries
we have developed a simplified model of intruder scattering. The main
reason why we resort to studying a simplified problem is the
difficulty in visualizing the dynamics and the structure of the
12-dimensional phase space in the full four-body Hill problem. We
stress, however, that, in the following, we are only {\it
interpreting} the dynamics exhibited in the full problem using
reduced dimensionality dynamics. None of the results obtained in
the previous section rely on this analysis.

Empirically it seems likely that the origin of the mass effect in
Figs.~\ref{fig3} and \ref{fig4} can be traced to the observed
difference in intruder dwell- or interplay-times \citep{shebalin}. This is because the only consistent difference we were able to find between simulations which were otherwise identical, except for having different mass ratios, was in intruder dwell-time distributions. A similar effect has also been
observed by Hills in simulations of star stellar-binary scattering
\citep{hills0, hills1, hills2, hills3}. However, Hills took the
binary orbits to be bound, two-body Keplerian orbits rather than
chaotic three-body orbits; {\it i.e.}, Hills' work does not contain
the equivalent of solar tides which, in our model, are essential for
the formation of quasi-bound binaries in the first place.

The model we build has its foundations in the following observations;

\begin{enumerate}

\item A separation in
timescales exists between fast intruder scattering and the timescale
of the mutual binary orbit. That is, chaos in the binary orbit
caused by solar tides generally develops on a slower timescale than
does the intruder scattering event.

\item The difference in dwell-times between equal and
unequal mass binaries is greatest for small intruders.

\item If simulations are done using zero-mass intruders then, clearly,
no binary stabilization can occur. Nevertheless, the intruder dwell-time
effect persists. In this limit the problem reduces to the so-called
restricted Hill four-body problem (RH4BP) \citep{sch}.

\item Experimental simulations reveal that the
dwell-time effect in the RH4BP exists even if the binary follows an
elliptical Keplerian orbit. These findings are based on extensive
simulations in the four-body Hill problem and in several variants of
the RH4BP in which different solutions for $\brho(t)$ were used --
see the discussion following equation (\ref{com}).

\item As the binary hardens then solar tides become relatively less important and the problem
reduces to the elliptic restricted three-body problem (ERTBP) \citep{sch}.

\end{enumerate}

\subsection{The Elliptic Restricted Three-body Problem}

Combining the observations in the previous subsection leads to the
following set of assumptions: neglect solar tides, assume elliptical
binary orbits and set the intruder mass to zero. After making these
approximations the four-body Hill problem reduces to the elliptic
restricted three-body problem \citep{sch} for which the Hamiltonian
is given by \citep{sze,llibre,mikkola, af};

\[
H_{e}=\frac{1}{2}((p_{x}+y)^{2}+(p_{y}-x)^{2}+{p_{z}}^{2}+z^{2})-
\]%
\[
\biggl(\frac{1-\mu^{\prime } }{\sqrt{(1+x)^{2}+y^{2}+z^{2}}}+\frac{\mu^{\prime } }{\sqrt{%
x^{2}+y^{2}+z^{2}}}+
\]%
\begin{equation}
\frac{1}{2}(x^{2}+y^{2}+z^{2})+(1-\mu^{\prime } )x+\frac{1}{2}(1-\mu^{\prime } )\biggr)\bigg/%
(1+e\,\cos \,f).  \label{ertbp}
\end{equation}

\noindent Here the semimajor axis of the {\it mutual} binary orbit
$a$ has been scaled to unity; $e$ is the eccentricity of the binary
orbit and $\mu^{\prime}=m_{2}/(m_{1}+m_{2})$ or $m_r =
\mu^{\prime}/(1-\mu^{\prime})$. The true anomaly of the binary
mutual orbit $f$, {\it i.e.}, the angular position measured from the
pericentre, is related to the physical time $t$ through \citep{sze}
\begin{equation}
\frac{df}{dt}=\frac{(1+e\,\cos \,f)^{2}}{(1-e^{2})^{3/2}}.  \label{true}
\end{equation}

The CRTBP Hamiltonian
\citep{cac}

\[
H_c=\frac{1}{2}({p_x}^2 + {p_y}^2 + {p_z}^2)-(x\,p_{y}-y\,p_{x})-
\]
\[
\frac{1-\mu^{\prime } }{\sqrt{(1+x)^{2}+y^{2}+z^{2}}}-\frac{\mu^{\prime } }{\sqrt{%
x^{2}+y^{2}+z^{2}}}-
\]
\begin{equation}
(1-\mu^{\prime })x - \frac{1}{2}(1-\mu^{\prime })  \label{crtbp}
\end{equation}

\noindent is recovered when $e=0$. The Hamiltonians (\ref{ertbp})
and (\ref{crtbp}) are defined in the rotating frame with the origin
transformed to one of the primaries.

Note that, in the following, the eccentricity being referred to in
the context of the ERTBP is the eccentricity of the {\it mutual} orbit
of the binary partners and not the {\it heliocentric} eccentricity
of the binary centre-of-mass. The binary centre-of-mass is still
assumed to be following a circular heliocentric orbit.

It is important to keep in mind that, in the ERTBP, the two
primaries are assumed to be following an elliptical (Keplerian)
mutual orbit and, therefore, these bodies cannot ionize. In our simplified model the ERTBP primaries are
actually the binary partners which are, in reality, following an
unbound chaotic orbit. However, if the intruder dwell-time is
sufficiently small then, {\it during the intruder scattering event itself},
neither the intruder nor the binary will sense the longer timescale
chaos caused by solar perturbations on the binary
orbit. In a sense this picture is similar to the Born-Oppenheimer
approximation for molecules in which the slow nuclei appear almost frozen
from the point of view of the much faster electrons \citep{BO}. It
should be emphasized that, even if the intruder does not sense that
the binary orbit is chaotic, the perturbation the intruder induces
on the binary can, nevertheless, influence the binary orbit's long
time development. That is the intruder can stabilize, destabilize,
or leave essentially unaffected the evolution of the binary mutual
orbit.

The ERTBP will, therefore, only be a reasonable model if the
timescale of the binary partner relative motion is much slower than
that of binary-intruder scattering encounters. If this is the case
then the four-body Hill problem can be approximately decomposed into
two simpler three-body systems; Sun-binary and binary-intruder. In
practice, we find that intruder dwell times -- the time the intruder
spends inside the Hill sphere -- in the full four-body simulations
are normally no more than a few percent of the Hill lifetime of the
orbit ($T_H$, defined earlier) and so we expect that studying the
reduced (ERTBP) dynamics will be useful as a proxy for the full problem.

Stated somewhat differently, we make the assumption that the
relative approach velocity of the massive binary partners is quite
small whereas hardening requires that the intruder approach in a
hyperbolic flyby at a somewhat larger relative velocity. While this
might seem contradictory, equipartition of energy in a particle disk
implies that the largest bodies will have the most circular coplanar
orbits, while the smaller bodies - here the intruders - will be more
eccentric and inclined. Thus the relative velocity of approach
should indeed be larger for the small intruders \citep{gold,
wetherhill, stewc, stewb, ohtsuki, ann}.

Because the
 binary is actually following an unbound chaotic orbit it is easily
disrupted. Even small intruders can, in principle, lead to early
disruption of the binary (as compared to $T_H$, the binary lifetime
in the absence of intruders). If, on the one hand, the intruder has a short
dwell-time then it is in and out of the Hill sphere quickly and
there is an opportunity for sudden energy transfer with minimal
disruption of the binary. Granted, this energy transfer may
destabilize the binary but, in the intruder velocity regime of
interest, we find that, on average, this is not the case -- again
this result is quite similar to the findings of Hills \citep{hills0,
hills1, hills2, hills3}. Note especially that ionization of the
binary is possible -- though not necessarily equally likely -- for
small as well as large intruders. This situation should be
contrasted with the case of already bound (Keplerian) binaries when,
in general, a comparably massive intruder may be needed to produce
complete or ``democratic'' ionization \citep{heggie}.

On the other hand if the intruder gets trapped in a relatively long-living
resonance with the primaries then there exists a greater opportunity
for disruption since the perturbation to the binary orbit is being
applied for longer. That is, if the intruder gets caught up in a
sticky chaotic layer its stay within the Hill sphere will be
extended (in a resonance) as compared to if it had entered an
asymptotically hyperbolic regime. While the lifetime of the
resonance may still be much shorter than $T_H$ it will be
significantly longer than the dwell time of an intruder undergoing
hyperbolic scattering.    Such resonances have three main decay
channels \citep{hills2, heggie}; (i) complete ionization, (ii)
expulsion of the intruder and (iii) expulsion of one of the original
binary partners (exchange). While channel (ii) is the most likely of the three for small intruders,
channels (i) and (iii) can also occur because the binary orbit is not bound. Thus, as compared to sudden
scattering, resonances (in the sense of long-lived trinary complexes
\citep{heggie}) tend, on average, to reduce the capture probability.

In view of the foregoing we suggest that the relative sizes of the {\it
chaotic} regions as compared to the {\it hyperbolic} regions will
correlate directly with the intruder's dwell-time. We argue that the relative
sizes of these regions should, therefore, also correlate to capture probabilities.
Chaos delays the intruder within the Hill sphere and so amplifies
the effect of the intruder on the binary orbit. Repeated
interactions -- and energy transfer -- between the binary and the
intruder tend to destabilize the fragile binary orbit. Importantly,
we are here comparing relative rather than absolute propensities for
stabilization. Not all fast encounters are stabilizing nor do all
resonances cause the binary to decay.

Thus, what requires examination is the relative size of {\it
chaotic} versus {\it hyperbolic} regimes for various mass
ratio/eccentricity combinations. The value of the ERTBP limit of the
four-body Hill problem is that it allows one to visualize phase
space directly; in this case using the Fast Lyapunov Indicator (FLI) as will now
be described (\citealt*{fli}; \citealt{af}).

\subsection{Fast Lyapunov Indicator}
The FLI is useful for mapping chaotic and regular regions of phase
space when Poincar\'{e} SOS cannot readily be computed
(\citealt{fli}; \citealt*{pfd}; \citealt{af}). Given an
$n$-dimensional continuous-time dynamical system,
\begin{equation}
{d\mathbf{x}}/{dt}=\mathbf{F}(\mathbf{x},t),\mathbf{x}%
=(x_{1},x_{2},...,x_{n}),  \label{EOM}
\end{equation}
the Fast Lyapunov Indicator is defined as \citep{fli}

\begin{equation}
FLI(\mathbf{x}(0),\mathbf{v}(0),t)=\ln |\mathbf{v}(t)|,  \label{fli}
\end{equation}

\noindent where \textbf{v}(t) is a solution of the system of
variational equations \citep*{tancredi}

\begin{equation}
\frac{d\mathbf{v}}{dt}=\biggl(\frac{\partial \mathbf{F}}{\partial \mathbf{x}}%
\biggr)\mathbf{v}.  \label{variation}
\end{equation}

Regularization \citep{KS, aarseth} was used to integrate equations
(\ref{EOM}) and (\ref{variation}).

\begin{figure*}
\begin{center}
\includegraphics{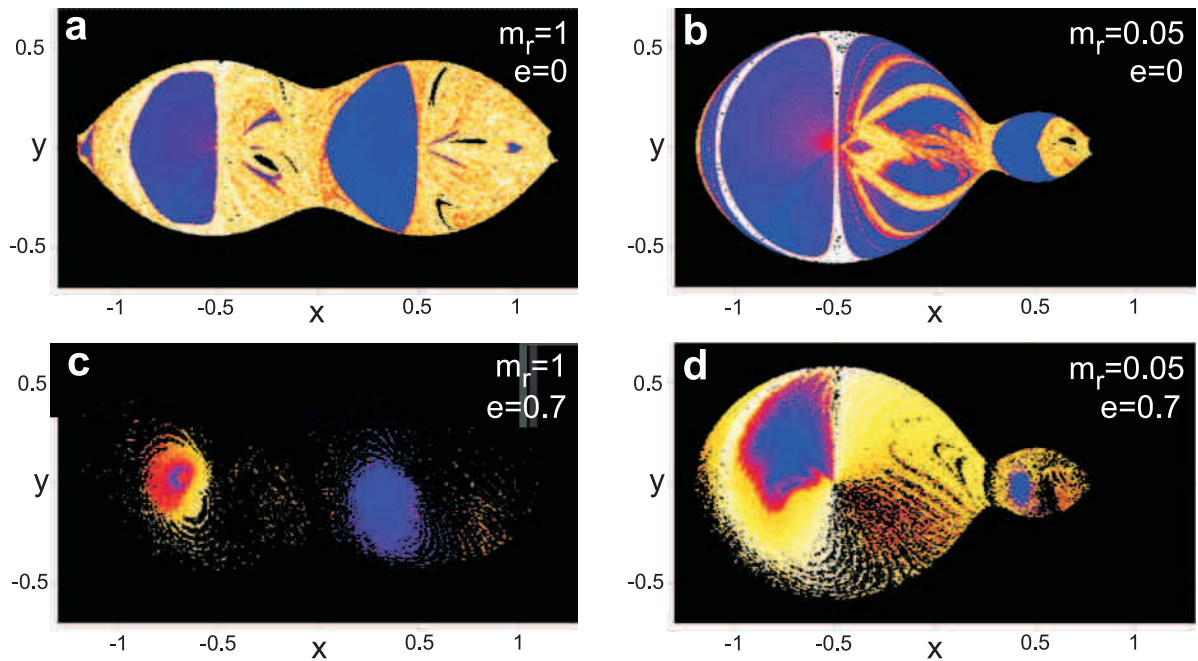}
\caption{\label{fig6} Fast Lyapunov indicator maps for the planar
ERTBP and four eccentricity/mass ratio combinations. The origin is
shifted so that the binary partners are located at $(x,y)=(\pm
0.5,0)$ in scaled units and here units are rescaled so that $R_H=1$.
Colour scale runs from blue to red to white with increasing FLI.
Initial energies are as defined in the text. Blank regions
correspond to directly scattering or hyperbolic trajectories which
survive for less than $\sim 20$ binary periods. Areas that resemble
shotgun-like patterns of points correspond to chaotic regions
\citep{af} (see the text). At the higher eccentricities scattering
(hyperbolic) regions are noticeably larger for (c) (equal masses)
than for (d) (unequal masses); this translates into enhanced capture
probabilities for equal masses.}
\end{center}
\end{figure*}

\subsubsection{Choice of initial conditions and energies in FLI maps}

To analyze the structure of phase space in our simplified system,
{\it i.e.}, in the ERTBP, we computed FLI maps in the planar (2D)
circular and elliptical cases. In the circular limit a direct
comparison with surfaces of section (in the Hill limit, $\mu^{\prime
} \ll 1$, \citealt{af}) can be made. We have previously demonstrated
that in the planar CRTBP the FLI
reproduces correctly all relevant features visible in the SOS
\citep{af}. In particular, the chaotic layer, as it evolves with
increasing energy, can be easily identified by high values of the
FLI. Note also that FLI measurements make sense even for relatively
short time integrations, and, even in 2D, are much more efficient
than is the construction of SOS \citep{fli, pfd, af}.

In the planar limit of the ERTBP initial conditions were generated
randomly within the Hill radius on the surface of section (see
Fig.~\ref{fig2} caption) taking, initially, $f = \pi/2$. This
guarantees that the ERTBP initial conditions reduce to those of
CRTBP \citep{af}. For a given $\mu^{\prime }$ and $e$ all ERTBP
initial conditions are, therefore, generated with identical initial
energies \citep{af}.

A further issue in generating the FLI maps in Fig.~\ref{fig6}
concerns the choice of initial energy. Because we are comparing
different masses and ellipticities it is not possible to work at
exactly the same energy in each case. This is because in the ERTBP
the energies of the Lagrange points, for example, depend on
ellipticity, mass and the instantaneous value of the true anomaly,
$f$. That is, energy (or the Jacobi constant, \citealt{murray}) is
not conserved in the ERTBP. Thus, in order to make a fair comparison
of the dynamics between these different cases we need to construct a
criterion of comparability; in practice, we selected initial values
of the Jacobi constant such that the sizes of the gateways in the
zero-velocity surface at time $t = 0$ (see Fig.~\ref{fig1}) were
comparable \citep{af}. Approximately, this equates the fluxes of
incoming/outgoing particles and facilitates a fair comparison of
relative densities of chaotic versus hyperbolic orbits. The actual
energies used at $t = 0$ ($f = \pi/2$) were $E = -1.725$ for $m_r =
1$ and  $E = -1.675$ for $m_r = 0.05$. Admittedly this is an {\it ad
hoc} approach but we find that the general appearance of the FLI
plots is quite robust to the precise values of the initial Jacobi
constant used provided that the gateways have similar sizes.

\subsubsection{Interpretation of FLI maps}

Figure~\ref{fig6} shows FLI maps for different mass ratios and
eccentricities. It is easy to distinguish between chaotic regions,
directly scattering (hyperbolic) regions and -- to the intruder --
inaccessible KAM regions \citep{af}. In the chaotic regions nearby
orbits tend to be dense and have similarly large FLI values. This
results in chaotic regions having the appearance of a shotgun-like
pattern -- unlike in a SOS, however, each point corresponds to the
initial condition of a {\it single} orbit coloured according to the
final computed value of its FLI. By contrast, in hyperbolic regions
orbits rapidly leave the Hill sphere and enter heliocentric orbits.
We chose 20 mutual orbital periods of the primaries (in the ERTBP
limit) as a cut-off such that orbits which survived for less than
this cut-off were not plotted. We verified that our results are not
sensitive to the precise choice of cut-off time. The large solid
regions in Fig.~\ref{fig6} correspond to regular (KAM) regions.
Elsewhere we have examined the effect of ellipticity on
chaos-assisted capture and the
 differences in FLI maps for chaotic, hyperbolic and
regular zones was examined in detail in the ERTBP \citep{af}.

In summary, in FLI maps hyperbolic regions manifest
themselves by their relative sparsity of points as compared to
chaotic and regular zones \citep{af}.

\subsubsection {Effect of mass ratio and eccentricity on capture probabilities}

First we compare the situation for equal versus unequal masses in
the case of circular orbits, {\it i.e.}, Fig.~\ref{fig6}a and
~\ref{fig6}b. Clearly the observable sea of chaos is much larger for
frame (a) {\it i.e.,} equal masses, than for frame (b) {\it i.e.,}
unequal masses. However, in both cases it is also true that there
are relatively few hyperbolic regions. For the case of unequal
masses the regular KAM regions are very large but since KAM regions
represent barriers in phase space \citep{ll} these regions are
inaccessible to the intruder. Thus in both Fig.~\ref{fig6}a and
~\ref{fig6}b an intruder is likely to become caught up in a chaotic
zone thereby producing a resonance. According to our earlier
arguments this will, on average, increase dwell-times and so tend to
destabilize the binary. This suggests that circular (or very low
eccentricity) binary orbits will be less likely to be captured.

Next we examine the case of higher eccentricity, $e = 0.7$, shown in
Fig.~\ref{fig6}c and ~\ref{fig6}d.  This case is actually the most
relevant to the real situation since the osculating eccentricity of
the chaotic binary is generally quite high. For equal masses, shown
in frame (c), very large hyperbolic regions are clearly visible
whereas in frame (d) (unequal masses) the ``non-regular" regions are
primarily chaotic rather than hyperbolic. Incoming intruders will
therefore encounter very different environments between the two
cases. If the binary partners are of equal mass, intruders will most
likely enter hyperbolic regions and so will exit the Hill sphere
promptly. In the case of unequal masses the battleground is of a
very different nature, resembling a field of sticky mud in which
intruders rapidly become bogged down. That is, they become
temporarily trapped in trinary resonances which increases the
probability that the binary will be disrupted rather than
stabilized. This can be explained, in part, by noting that in the
CRTBP with equal mass primaries (the Copenhagen problem,
\citealt{sze, benet, nag}), the zero velocity surface is symmetric
in $x$ and so the gateways into and out of the capture region are
equivalent. This facilitates quick transits through the capture
zone, whereas in the asymmetric mass system one gateway is smaller
than the other constituting a bottleneck which hinders passage.

The FLI maps in Fig.~\ref{fig6} provide a qualitative explanation of
the observed mass effect. The dwell-times of intruders within the
Hill sphere are a sensitive function of whether intruders enter
hyperbolic or chaotic regions of phase space inside the Hill sphere.
As system parameters change (total initial energy, ellipticity,
etc.) the sizes and shapes of these regions alter in a highly
nonlinear manner (see Fig.~\ref{fig2}c) as is characteristic of
chaotic scattering. With increasing ellipticity the sizes of the
chaotic versus the directly scattering regions {\it decrease},
especially in the case of equal mass binary partners. This
translates into enhanced capture probabilities for equal masses.

Overall, based on a comparison of the four cases in Fig.~\ref{fig6}
in the ERTBP, we conclude that intruder stabilization will be most
efficient for (i) equal masses and (ii) moderate to high
eccentricity orbits. It is somewhat ironic that, in this mechanism,
the existence of chaos in the binary orbit (produced by solar tides)
is critical to nascent KBB formation, whereas if intruders enter
zones of chaotic (as opposed to hyperbolic) scattering then the
result is destabilization of the binary.

However, the ERTBP model has its limitations. The assumption that
the mutual chaotic binary orbit can be treated (from the point of
view of the intruder) as an elliptical Kepler orbit on short
timescales is not valid if the actual binary orbit is itself
extremely unstable on comparable timescales. Such three-body orbits tend to involve ``near misses'' of the
binary partners and are less stable than are more ``circular'' three-body orbits
({\it e.g.,} see Fig.~\ref{fig2}a). Thus for very high instantaneous
eccentricities (orbital segments involving near misses) the binary
partners (in the full four-body problem) are difficult to stabilize. This
tends to reduce the capture probability into very high eccentricity
orbits considerably.

The net effect of all of these considerations is to favor moderately
elliptical binary mutual orbits when the binary partners have
approximately equal masses. This is precisely what is found in our
simulations and in actual observations of KBBs
\citep{burns,noll1,noll2}.

\section{Alternative Formation Models}
\label{sec5}

 Of the four paths mentioned in the Introduction and summarized in
Table~\ref{tbl} extensive numerical simulations have only been
reported for the exchange mechanism, Path~4 \citep{funato, kol}. In
this section we present a brief comparison of our multiple
scattering model with alternative theories and also report some
model simulations which we have performed for Paths 2 and 3
\citep{gold}.

\subsection{Exchange reactions}

In the exchange reaction model \citep{funato} the first step is the
formation of a binary through a two-body dissipative encounter. This
can happen in two ways; (i) tidal disruption of an object during a
close encounter followed by accretion of the resulting fragments to
produce a binary; and, (ii) a ``giant'' impact in which coagulating
debris after the collision produces a satellite orbiting a larger
object. Main-belt asteroid binaries (and the Earth-Moon binary
\citep{canup, canupa} and (with some important differences) possibly the Pluto-Charon binary \citep{canup1})
are thought to have formed in this manner. Generally, the result is
a binary with an extreme mass ratio and a tight, circular orbit. In
the exchange reaction mechanism, the mass ratio of such
proto-binaries in the KB is increased through later
gravitational scattering encounters with a third object originating
(in the simulations) at infinity. \citet{funato} modelled this by
performing extensive scattering simulations between strongly
asymmetric-mass binaries already following bound, compact, circular
orbits and large incoming intruder masses. They found a propensity
for the smaller member of the initial binary to be replaced by the
intruder with the final binary then following a large, highly
eccentric Keplerian orbit.

While this mechanism is plausible and does produce binaries which
are at least qualitatively similar to known KBB orbits a number of
difficulties nevertheless remain. In the first place, this
mechanism does not provide an {\it ab initio} explanation for why the
the mass ratios of actual KBBs are of order unity. Instead, it simply provides a
route by which a binary can increase its mass ratio. Because Funato et al.
assumed that the primary member of the initial binary and the
intruder had equal masses then, if exchange occurs, a binary of
order unity mass ratio is guaranteed to result. There appears
to be no compelling reason, at least based on the simulations
reported \citep{funato}, to conclude that mass ratios of order
unity should necessarily result from this process, only that
exchange reactions can increase mass ratios.

In addition, exchange reactions seem to produce orbital
eccentricities (and semimajor axes) larger than those typically
observed \citep{noll,noll2}. However, this might simply be the
result of the particular choice of masses made by \citet{funato}.
Thus it would be interesting to extend these simulations to
include a broader range of intruder masses so as to understand (a)
the role of intruder mass in determining the probability of
exchange and (b) the effect of intruder mass on post-exchange
orbital properties of the binary.

\subsection{Two-body collisions inside the Hill sphere}

The first mechanism proposed for the production of trans-neptunian
binaries is that of \citet{weiden}. In this model two bodies
accrete after a mutual collision within the Hill sphere of a
third, larger, body. They then remain bound as a single object
orbiting the larger body, thereby producing a binary. The main
objections to this model have to do with the scarcity of large
objects in the primordial KB. However, \citet{petit} \citep[see
also][]{stern} discuss this issue at length and conclude that the
mechanism of \citet{weiden} might have operated, possibly in
tandem with other mechanisms. In the absence of detailed
simulations it remains an open question whether this mechanism can
provide quantitative agreement with the observed properties of
KBBs.

\subsection{Capture through dynamical friction}

This mechanism was originally proposed by \citet{gold}. Here we have
shown that the dynamical basis for this model is chaos-assisted capture
\citep{cac,af}. No detailed simulations of KBB formation or
their orbital and compositional properties were presented by
\citet{gold}. Because this mechanism proceeds from similar
assumptions to the one we propose, we performed a limited number of
simulations in which multiple intruder scattering as an energy loss
mechanism was replaced by dynamical friction, modelled as simple
velocity dependent dissipation \citep{murray,cac}. We find that,
 dynamical friction provides an efficient route to binary
capture and Keplerization, producing final orbital elements similar
to those in Fig.~\ref{fig4}. However, it is not sensitive to the mass ratio
of the initial quasi-bound binary. This is because the scaled Hill
equations (\ref{hill}), even including dissipation, contain no
parameters and so cannot depend on the masses of the binary
partners. Thus, no mass effect is predicted. Of course, more
sophisticated models of dynamical friction \citep{chan, bin, ann,
just} might conceivably reveal such a preference. In addition,
dynamical friction has been ruled out on the basis of estimates of
planetesimal mass distributions although this issue may be worth
revisiting \citep{funato}.

\subsection{The $L^{3}$ mechanism}

\begin{figure}
\begin{center}
\includegraphics{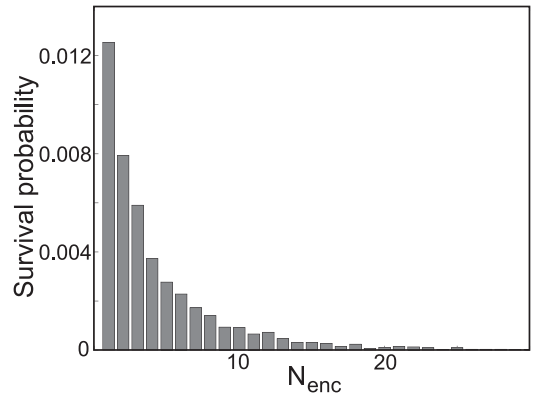}
\caption{\label{fig7} Histogram showing the survival probability of
equal mass chaotic binaries (${\alpha}_1={\alpha}_2$) after
$N_{enc}$ $L^{3}$ encounters with large intruders (${\alpha}_3=0.3$,
see equation (\ref{alpha})). 1000 initial binaries were taken with
initial conditions for three-body Hill orbits and intruders
originated as described in Sec.~\ref{sec3}. To generate statistics
the simulation was repeated 50 times, each time choosing the phase
of each binary randomly with respect to the intruder. }
\end{center}
\end{figure}

Again, this mechanism was originally proposed by \citet{gold} and is
the most similar to the one which we present here. Although no
simulations of $L^3$ were made by \citet{gold} a number of issues
can be raised in regard to it; (a) in $L^{3}$, as originally
proposed, roughly equal mass binaries result directly from the
assumption that a transient binary undergoes a scattering encounter
with a third body of comparable mass to the binary partners (which
are assumed to be themselves of similar mass). That is, the
preference for order unity mass ratios is an inescapable result of
the foundational assumptions of the model; (b) unless multiple
$L^{3}$ type encounters are being assumed, then hardening and
Keplerization of the large transitory binary must occur in a single
encounter which seems unlikely. To test this we have performed
simulations exactly as for Fig.~\ref{fig4} except using a large
intruder. The results are summarized in Fig.~\ref{fig7} and indicate
that multiple large body encounters tend to disrupt the binary
catastrophically. Further, examination of the actual orbital
elements produced in these simulations confirms that neither a
single nor a small number of $L^{3}$-type events is usually
sufficient to produce Keplerian orbits. Instead, after a small number
of such encounters the orbits resemble those of the initial
quasi-bound binary, {\it e.g.}, that shown in Fig.~\ref{fig2}. In
these cases Keplerian orbital elements cannot meaningfully be
defined.

Of course a hybrid of our model and $L^3$ is also possible. Certainly our
simulations do not rule out the possibility of $L^3$-type collisions
with additional hardening being produced through collisions with
smaller bodies; however, $L^3$  collisions are not necessary for
capture or to produce roughly equal mass binaries.

\section{Conclusions}
\label{sec6}

We have presented what are, to our knowledge, the first simulations
of binary-intruder scattering in which the binary orbit is itself
chaotic. Previous simulations of binary star intruder scattering
have focussed on the important but simpler case in which the initial binary already follows a
bound Keplerian ellipse. Here the binary orbits used are long living
chaotic orbits (in the Hill approximation) which cling for long
periods to regular KAM structures in phase space. These transitory
orbits are possible only if the Sun-binary Hill sphere is taken into
account. The importance of the Hill sphere in temporarily capturing
Kuiper-belt binaries was first noted by \citet{gold} on empirical
grounds. The dynamical explanation for the results proposed here is chaos-assisted
capture \citep{cac,af,holman}.

Once a proto-binary has formed within the Hill sphere then, if it is
to survive, some stabilization mechanism must operate. Our
simulations demonstrated that multiple low-mass intruder scattering
can not only stabilize transitory binaries but also provides an
efficient route to binary hardening and Keplerization of the orbit.
By adjusting a single parameter -- the number of intruder-binary
encounters -- we were able to obtain excellent agreement with the
orbital properties of known KBBs. In particular, our simulations
predict that KBBs will have moderate eccentricities. This is in good
agreement with recent observations \citep{noll,noll1,noll2} and
contrasts with the exchange mechanism of KBB formation which
produces exclusively large eccentricities \citep{funato, kol}.

Our simulations also lead to a striking preference for the
production of binaries having roughly equal masses. Strongly
asymmetric mass ratios -- typical of main belt asteroid binaries --
are selected against, and, again this appears to be consistent with
observations. The model we propose is not set up to produce only
order unity mass ratio binaries; nevertheless, a preference for mass
ratios of order unity emerges.

An important difference
between our model and $L^3$ concerns how the binary hardens.
Single encounters with large intruders are unlikely to produce the
relatively compact (as compared to $R_H$) Keplerian binary orbits
that are actually observed. Moreover, we find that multiple
encounters with large intruders tend, on average, to disrupt the
binary. Instead, the model we propose involves a sequence of
low-mass intruder events rather than a single scattering encounter
with a massive object. It turns out that this provides an
efficient mechanism for binary hardening and Keplerization.
However, because we are unable to estimate the
probabilities for the initial rate of production of quasi-bound
binaries of varying mass ratios in the Hill sphere, it is unclear
how strongly this mass effect will persist. Nevertheless, our
simulations demonstrated that equal mass binaries have a much
higher survival probability after multiple encounters than do
asymmetrical mass ratio objects. Thus, we are confident that the
mass effect we have found will not be washed out by the statistics
of transitory binary formation.

Throughout we have assumed that the binary centre-of-mass follows a
circular (heliocentric) orbit. However, in principle, we also need to consider the
effect of introducing eccentricity into the heliocentric binary
orbits. This is because some observed KBBs have significantly eccentric
heliocentric orbits with perihelia close to, or even inside,
Neptune's orbit ({\it e.g.,} Pluto-Charon, $(47171)~1999~TC_{36}$
and $(26308)~1998~SM_{165}$). Typical KBB heliocentric ellipticities
lie in the range $\sim 0.03 - 0.37$ \citep{noll}. In our recent
study of capture and escape in the elliptic restricted three-body
problem \citep{af} we found that the introduction of moderate
heliocentric ellipticity reduces, but does not entirely eliminate, the
efficacy of the chaos-assisted capture mechanism. While further study of this issue is
clearly warranted, we expect that the results presented here will similarly
carry over to the elliptic heliocentric case.

Of course, it is also possible that observed heliocentric TNO binary
eccentricities are not primordial. Actually, this is suggested by the
observation that massive bodies in eccentric orbits will tend to
have large approach velocities apart from coincidences such as, {\it e.g.},
when perihelia are aligned. Because such cases are rare, relative velocities will tend to be large and this will serve to reduce the
probability of capture into chaotic orbits. This conclusion is consistent with
results for capture in the elliptic restricted three-body problem
\citep{af}. These considerations suggest that, as with Paths 1 - 3,
binaries formed when the disk was massive and dynamically cold, {\it
i.e.,} prior to the dynamical excitation and depletion of the
Kuiper-belt \citep{dunc, gomes, lev}.

Finally, we note that Figs.~\ref{fig3}-\ref{fig5} demonstrate that,
while significant, the selection for order unity mass ratio objects
is not to the complete exclusion of smaller mass ratios. For
example, the differences in dwell-time distributions are not so
great as to eliminate the capture of small mass-ratio binaries
entirely. Rather, Figs.~\ref{fig3} and \ref{fig4} lead us to predict
that a sizeable population of asymmetric mass KBBs is probably
awaiting detection. Heralds of this group of objects might already
exist in the low mass ratio binary objects $(47171)~1999~TC_{36}$
\citep{iau1,alt} and $(26308)~1998~SM_{165}$
\citep{iau2,stern,noll,petit}.

\section*{acknowledgments}

This work was supported by grants from the US National Science
Foundation through grant 0202185 and the Petroleum Research Fund
administered by the American Chemical Society. All opinions
expressed in this article are those of the authors and do not
necessarily reflect those of the National Science Foundation.
S.~A.~A. also acknowledges support from Forschungszentrum
J\"ulich. We thank Kevin Hestir for valuable suggestions and
comments on the manuscript. We also thank two anonymous referees for
helpful and insightful suggestions.

\end{document}